\documentclass{svproc}
\usepackage{url}

\usepackage{epsf,color,colordvi}
\usepackage{caption}
\usepackage{graphicx}
\usepackage{amsmath}
\usepackage{times}
\usepackage{subfig}
\usepackage{ulem}  
\usepackage{placeins}

\begin{document}
\mainmatter              
\title{Analyzing the mesonic spectrum using the method of Schottky anomaly }
\titlerunning{Schottky Anomaly}  
%
\author{Aritra Biswas\inst{1} 
}
%
%
%
\institute{The Institute of Mathematical Sciences, Chennai 600113, India (affiliated to HBNI),\\
\email{aritrab@imsc.res.in},\\ WWW home page:
\texttt{https://sites.google.com/site/iluvnpur/contact}
}

\maketitle              

\begin{abstract}
A possible diagnostic is proposed which may be used to infer the different scales underlying the dynamical structure of hadronic resonances using
the phenomenon of Schottky anomaly.
\end{abstract}
\section{Introduction}
The search for exotic hadrons has been rejuvenated in recent years with the discovery of the so called ``pentaquark states''~\cite{pentaquark}. Along with the X, Y, Z states~\cite{Olsen:2014} discovered in the last decade by the different experimental groups~\cite{olsen_ref} and the recent confirmation of the $\Lambda(1405)$ as a molecular state~\cite{lattice} (which had been proposed earlier~\cite{rajaji1}), this has rekindled the quest for a better theoretical understanding of these states. Various models have been proposed to this end over decades~\cite{Godfrey:olsen}. We propose a model independent analysis in order to identify states which differ in their underlying dynamics due to different interaction scales responsible for forming composite hadronic states. The tool used is the method of Schottky anomaly. We restrict ourselves to the analysis of the mesonic spectra only.

\section{$\textbf{C}_{\textbf{V}}$ for an ideal system}
\begin{figure}
\centering
\subfloat[][]{\includegraphics[width=0.5\textwidth]{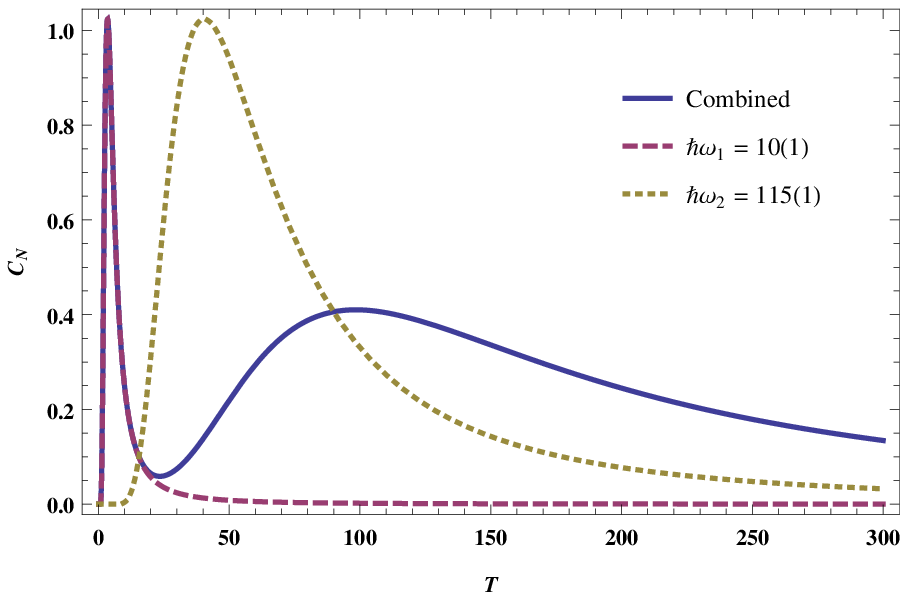}}\hfill
\subfloat[][]{\includegraphics[width=0.5\textwidth]{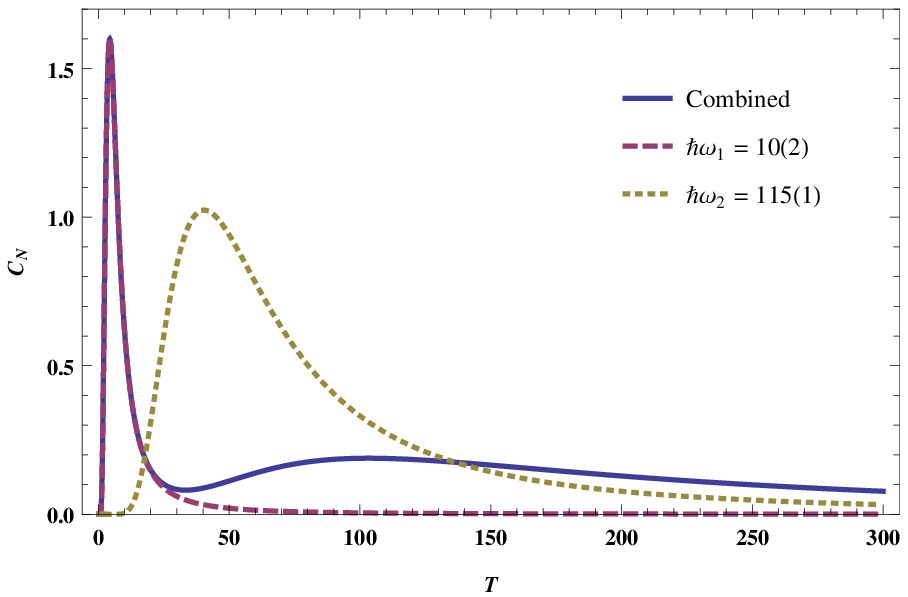}}\\
\caption{Schematic illustration of the Schottky peaks in the ideal case when the data
contains two scales. The individual Schottky peaks corresponding to separate spectra (harmonic oscillator
spectrum), as well as the combined one (solid line), are shown. The frequencies 
are the same for all of the four figures as shown ($\hbar\omega_1=10$ MeV and $\hbar\omega_2=115$ MeV).
The cutoff in the principle quantum number $n$ (indicated in brackets) are: (a) $n_1=n_2=1$ and (b) $n_1=2$
$n_2=1$.}
\label{fig1}
\end{figure}

Consider a two-level system with energy-gap $\Delta$. The canonical partition function is given by
\begin{equation}
 Z=1+e^{-\beta\Delta}.
\label{idealz}
\end{equation}
where $\beta=1/k_BT$ is the inverse temperature.\footnote{The temperature is introduced here simply as a mathematical parameter to
define the partition function of the system and no assumption is made regarding the system being
in a heat bath in equilibrium.} The specific heat of the system at constant volume may be defined as
\begin{equation}
C_V=\beta^2\left[\frac{1}{Z}\frac{\partial^2 Z}{\partial\beta^2}
-\left(\frac{1}{Z}\frac{\partial Z}{\partial\beta}\right)^2\right]
=\beta^2[\langle E^2\rangle-\langle E\rangle^2].
\label{cvdef}
\end{equation} 
Substituting
for the partition function of the two level system given in eq.(\ref{idealz})
we have
\begin{equation}
C_V= \beta^2 \frac{\Delta^2 e^{-\beta\Delta}}{(1+e^{-\beta\Delta})^2}.
\end{equation}
When $C_V$ is plotted against $\beta\Delta$ the Schottky peak appears at a value $\beta\Delta\approx 2.4$ . The location of the peak is a function of the energy gap in the system. A realistic spectra in general might have more than one scale in operation. In order to simulate the effect such a system we first consider the spectrum of a three-dimensional Harmonic oscillator. The partition function of the system is given by $Z_1=\sum_{n=0}^{\infty}D(n)e^{-\beta\hbar\omega(n+3/2)}$
where the oscillator parameter $\omega$ defines the scale in the problem and $D(n)$ is the degeneracy of the level. We combine the spectra of two such systems with different values for the scale parameters (Fig.~\ref{fig1}). The effect of combining is to normalize the specific heat with a single partition function given by,
\begin{equation}
Z(\beta)= Z_1(\beta,\hbar\omega_1)+Z_2(\beta,\hbar\omega_2).
\end{equation}
The peaks appear when there spectrum has a cut off 
otherwise, it will saturate as a function of temperature. In the next section we give our results for the analysis of the current heavy mesonic spectra taken from~\cite{pdg}. For a detailed discussion the interested reader is referred to~\cite{Schottky}.
\section{Analyzing the experimental spectra}
\begin{itemize}
\item\textbf{The charmonium spectra}:
When all the $c\bar{c}$ states are plotted together, the existence of two peaks at $T\approx40$ MeV and $T\approx190$ MeV are revealed, indicating the existence of two well defined scales, which might be identified with the ``hyperfine'' and the ``confinement'' scales respectively. When the states are grouped according to their individual $J$ values $J=0,1,2$, the hyperfine peak vanishes, and only the confinement peak is retained. However, when only the ``exotic'' states in the charmonium mass range are plotted, we again find a double-peaked structure, where the confinement peak shifts to $100$ MeV along with a sharp peak below $10$ MeV. 
\item\textbf{The bottomonium spectra}
The two peaked structure with distinct hyperfine ($22$ MeV) and confinement ($185$ MeV) peaks are revealed for the case of the bottomonium spectrum also. The hyperfine peak disappears as expected when the states are grouped according to their $J$ values and plotted again.The bottomonium exotics are however, scarce in number (three to be exact), with ill defined quantum numbers. It is hence premature to comment on these states. However, on plotting, they do retain the peak at $100$ MeV, which is encouraging and may mean that the underlying mechanism for the charmonium and the bottomonium exotic states are the same.
\item\textbf{The open charm spectra}
Compared to those discussed previously, this case is more complicated due to the presence of the isospin ($I$) symmetry along with the $J$ symmetry. However, on plotting all the open charm mesons together, the corresponding $C_V$ vs $T$ plot shows the two-peaked structure with the confinement peak at $T=160$ MeV and the hyperfine peak at $T=50$ MeV approximately. On plotting states with the same $J$ and $I$ together, the hyperfine peak vanishes and only the confinement peak is retained. 
\item\textbf{The open bottom spectra}
The open bottom spectra is rather sparse as compared to the open charm scenario. Nevertheless, some features are already visible. The two peaked structure prevails on plotting all the states together. On grouping the states according to their $J$ and $I$ values, a well defined confinement peak is seen. However, due to the limited number of states, one should wait for the advent of more data in the future before drawing any conclusions for this sector.
\end{itemize}

%
%
\end{document}